\documentclass[aps, pra, reprint, amsmath, amsfonts, amssymb, superscriptaddress]{revtex4-1}
\usepackage{rotating}
\usepackage{graphicx}

\def\KTH{KTH Royal Institute of Technology, Nanostructure Physics, Albanova, SE-10791 Stockholm, Sweden}

\def\KTHSURF{KTH Royal Institute of Technology, School of Chemical Science and Engineering, Department of Chemistry, Surface and Corrosion Science, Drottning Kristinas v\"ag 51, SE-100 44 Stockholm, Sweden}

\def\MONS{Laboratory for Chemistry of Novel Materials, Center for Innovation and Research in Materials and Polymers, University of Mons (UMONS), Place du Parc 20, B-7000 Mons, Belgium}

\def\UChicago{James Franck Institute, University of Chicago, 929 East 57th St., Chicago, IL 60637, USA}

\begin{document}

\title{On modeling and measuring viscoelasticity \\ with dynamic Atomic Force Microscopy}

    \author{Per-Anders Thor\'en}
    \affiliation{\KTH}

    \author{Riccardo Borgani}
    \affiliation{\KTH}
    
    \author{Daniel Forchheimer}
    \affiliation{\KTH}

    \author{Illia Dobryden}
    \affiliation{\KTHSURF}

    \author{Per M. Claesson}
    \affiliation{\KTHSURF}

    \author{Hailu G. Kassa}
    \affiliation{\MONS}

    \author{Philippe Lecl\`ere}
    \affiliation{\MONS}
    
    \author{Yifan Wang}
    \affiliation{\UChicago}
    
    \author{Heinrich M. Jaeger}
    \affiliation{\UChicago}

    \author{David B. Haviland}\email{haviland@kth.se}
    \affiliation{\KTH}

\begin{abstract}
The interaction between a rapidly oscillating atomic force microscope tip and a soft material surface is described using both elastic and viscous forces with a moving surface model.  We derive the simplest form of this model, motivating it as a way to capture the impact dynamics of the tip and sample with an interaction consisting of two components:  interfacial or surface force, and bulk or volumetric force.  Analytic solutions to the piece-wise linear model identify characteristic time constants, providing a physical explanation of the hysteresis observed in the measured dynamic force quadrature curves.  Numerical simulation is used to fit the model to experimental data and excellent agreement is  found with a variety of different samples.   The model parameters form a dimensionless impact-rheology factor, giving a quantitative  physical number to characterize a viscoelastic surface that does not depend on the tip shape or cantilever frequency.
\end{abstract}

\maketitle

\section{Introduction}

An increasingly important application of the Atomic Force Microscope (AFM) is the characterization of viscoelastic materials and interfaces, such as cell membranes and tendons\cite{Radmacher_viscoelastic_creep_cells_2016,Efremov:NanoscaleViscoelastic:2017,Hecht:ViscoelasticAFMcells:2015,Connizzo:TendonHighFrequencyAFM:2017}, polymer blends and composites \cite{Hurley:DynamicContactMethods:2013,Stan:IntermittentContactAFM:2014,Yablon:TempertureDependantLossTangent:2014,Tsukruk:NanomechanicalAFMreview:2015,Nakajima:ViscoelasticPolymersBimodal:2016,Turner:LossTangentPolymers:2017,Claesson:PCCP:2017}, the liquid-gas and liquid-solid interfaces \cite{Kilgore:ViscoelasticLiquidSolidInterface:2015,Maali:ViscoelasticDragAirWaterInterface:2017},  and suspended membranes \cite{Schlicke:AFMbuldgeTest:2014}. The present trend toward higher scanning speeds \cite{Ando:HighSpeedAFMreview:2013,Miles:HighSpeedAFMreview:2013} and higher resolution mapping of mechanical properties \cite{Garcia:ElasticMappingLiquid:2017,Raman:FastMechanicalMappingCells:2015} necessitates more rapid tip motion and therefore a larger viscous contribution to the tip-sample force.   Viscous force may even dominate over elastic force with soft materials and a proper characterization of the material must therefore rely on a dynamic measurement that distinguishes viscous from elastic force.  Here we employ such a method to demonstrate how traditional models for extracting material properties from AFM data fail to explain measurements on soft materials.  We introduce the most simple form of an alternative model that does explain the data.  We show how our model parameters relate to physically meaningful numbers that characterize mechanical response of the viscoelastic surface.

The AFM tip oscillating in and out of contact with the sample is a nanometer-scale example of a broader class of dynamical systems known as impact oscillators.  These are often modeled with piece-wise non-smooth impact forces that produce interesting bifurcations \cite{Peterka1999}. The oscillation in dynamic AFM typically has a frequency close to a cantilever resonance with high quality factor $Q$, where the inertial and linear restoring forces of the cantilever body dominate the system dynamics. Nevertheless, much smaller impact forces that weakly perturb the harmonic motion of the cantilever can be obtained with the help of lockin measurement techniques.  We use a well established multifrequency lockin method where many Fourier components of this nonlinear perturbation are detected as intermodulation products near resonance~\cite{Platz:ImAFM:08},  providing a great deal of information about the impact forces.

When the rigid AFM tip impacts the soft sample it experiences different types of force:  elastic force associated with strain in the contact volume and curvature of the contacting interface, and viscous force associated with the rate of change of strain and curvature.  These sample deformations occur at the nanometer scale, where the surface-to-volume ratio is much larger than that of macroscopic contact mechanics.  We therefore expect interfacial forces to play an important role in AFM.  A viscoelastic model of AFM contact forces must look beyond bulk rheology to also include the rheology of the interface\cite{Miller:InterfacialRheology:2010}.

\begin{figure}[ht]
\begin{center}
\includegraphics[width=7cm]{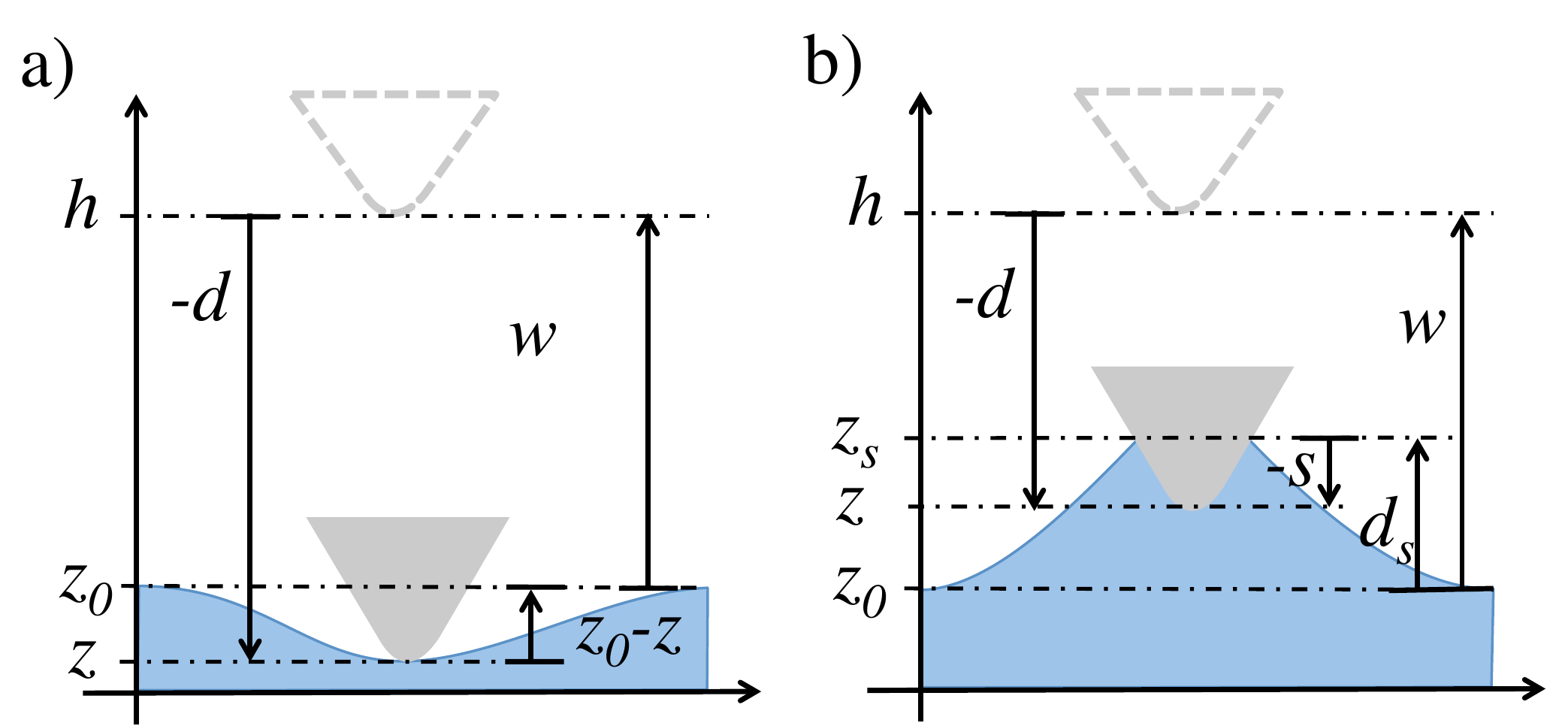}
\includegraphics[width=7cm]{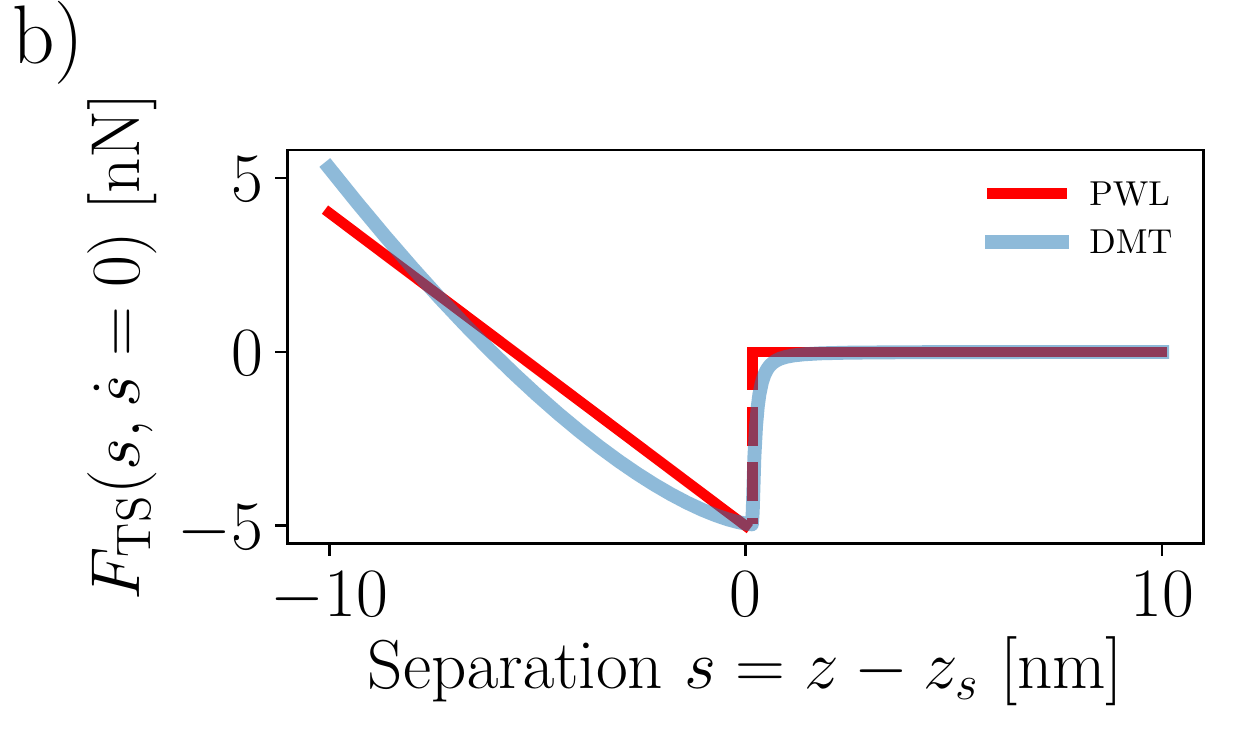}
\caption{{\bf Coordinates and piece-wise-linear (PWL) interaction}. The cantilever deflection $d=z-h$ is measured by the AFM detector, where $z(t)$ is the instantaneous tip position and the constant $h$ is the equilibrium (zero force) tip position.  $z_0$ is the equilibrium position of the sample surface.  (\textbf{a}) The traditional view of contact forces in AFM has the tip and surface moving together when they are in contact, $z=z_s$, and interaction force is considered to be a function of the surface indentation $(z_0-z)$.  (\textbf{b}) The moving surface model treats the surface position $z_s(t)$ as an independent dynamic variable.   The model introduces elastic and viscous forces that depend on surface deflection, $d_s=z_s-z_0$ and velocity $\dot{d_s}$, and an interaction force that depends on the separation $s=z-z_s$ and $\dot{s}$ .  Forces are balanced in the inertial reference frame where the cantilever has a fixed working distance to the sample $w=h-z_0$.  (\textbf{c}) The piece-wise linear (PWL) interaction force plotted together with a calculated DMT model frequently used in AFM~\cite{melcher:061301}.  The parameters for the DMT model are:  reduced modulus $E^*=100$~MPa, Hamaker constant $H=8\times10^{-20}$~J, tip radius $R=10$~nm, inter-molecular spacing $a_0=0.16$~nm.  Parameters for the PWL model are:  $F_\text{ad} = 5$~nN, $ k_\text{v}=0.9$~N/m.  The range of separation shown is that typically covered by the cantilever oscillation for dynamic AFM on soft materials.}
\label{fig:fts}
\end{center}
\end{figure}

The traditional approach to quantitative AFM is based on contact mechanics, where tip-sample force is considered to result from bulk elasticity.  The Hertz model~\cite{Hertz:contact:1881} gives a repulsive contact force.  Assuming an axially symmetric rigid tip, one can parameterize the tip profile in terms of a power $m \in [1,2]$ and a length scale $\ell$ \cite{Labuda2016}, to express the contact force as a function of tip indentation $z_0-z$, where $z$ is the position of the tip and $z_0$ is the equilibrium position of the surface (see fig.~\ref{fig:fts}a). 
\begin{equation}
F_\text{con}(z) =2 E_\text{eff} \alpha_c(m) \ell^{2-m} (z_0-z)^m 
\end{equation}
Here $E_\text{eff}$ is the effective modulus. Typically one does not know $\ell$, $m$, or $z_0$ and they should therefore be treated as free parameters when fitting the measured force curve $F_{TS}(z)$ to extract $E_\text{eff}$. With stiff samples one should not assume that the tip is rigid, in which case several more parameters are required to actually get the material modulus from $E_\text{eff}$.  Whatever the case, this class of models gives only constant forces that are not explained by dissipation.  Such models can not say anything about the sample viscosity, as it gives rise to a velocity-dependent force.  

A viscoelastic version of the Hertz model has been studied in the context of two-body collisions, where impact forces can be calculated in the center-of-mass reference frame~\cite{Wettlaufer:HertzBeyondBelief:2014}.   With the center-of-mass in the laboratory frame, defined as the inertial reference frame where the entire sample is at rest ({\em i.e.} sample mass $\gg$ cantilever mass), the viscoelastic Hertzian model reduces to a Kelvin-Voigt expression for the tip-sample force
\begin{equation}
F_\text{TS} = F_\text{con}(z) + \dot{z}\eta(z),
\end{equation}
where dissipation is introduced via a viscous damping coefficient $\eta(z)$ that depends on tip position.  One can in principle extract $\eta(z)$ from a dynamic AFM measurement~\cite{Platz:RoleNLDynamics:2012}, but relating it to sample viscosity would involve a complicated model requiring knowledge of the tip geometry.  
Other models of contact viscoelasticity use a creep-compliance picture, where the elastic modulus is time-dependent and force is determined by integrating over deformation history~\cite{Radmacher_viscoelastic_creep_cells_2016,Efremov:NanoscaleViscoelastic:2017,Hecht:ViscoelasticAFMcells:2015}.  One can also use finite element methods with linear~\cite{Solares:StandardLinearSolid:2016} or nonlinear force-displacement relations that account for an attractive tip-sample force~\cite{Attard:MeasurementInterpretationViscoelastic:2007}.  

Independent of these bulk viscoelastic models, tip-sample force in AFM may also result from interfacial energy or surface tension $\gamma$.  The work of adhesion is the change in total surface energy after contact,
\begin{equation}
    W= (\gamma_\text{T} + \gamma_\text{S}) - \gamma_\text{TS}
\end{equation}
where the subscripts $T$ and $S$ refer to tip and sample respectively.  Typically $W>0$ giving attractive force upon contact.  The role of surface energy in small mechanical contacts was originally discussed by Johnson, Kendal and Roberts (JKR)~\cite{JKR:71}.  The JKR model has been shown to break down when the contact radius of curvature $R$ is small in comparison with the elastocapillary length, $L= \Upsilon / E$, where $\Upsilon$ is the surface stress and $E$ is the Young's modulus of the sample bulk~\cite{Style:SolidSurfaceTension:2013, Butt:ForceBetweenStiffSoft:2017}.  For a soft material $E\sim$~3~MPa forming a contact with relatively low interfacial energy $\gamma_\text{TS} \sim$~30~mN/m~ and no additional surface stress (in which case  $\Upsilon=\gamma_\text{TS}$), we find $L\sim$~10~nm, the typical radius of an AFM tip.   

Thus, an AFM tip contacting a soft material should resemble a liquid-like sample wetting and forming a meniscus around the tip, as opposed to the tip compressing an elastic solid.  Indeed,  $W$ is called the spreading parameter in the context of wetting phenomena if the sample corresponds to the liquid and the tip to the solid substrate, and $W>0$ is called total wetting.  

The discussion above makes clear that a three-dimensional continuum model of AFM impact forces is difficult to formulate if one includes both bulk and interfacial phenomena, and both elastic and viscous contributions.  Knowledge of the tip and sample geometry is required, something that is difficult to determine either before or after an AFM measurement.  Even if a realistic model could be formulated, determining its many free parameters from AFM data appears to be a hopeless task because the AFM measurement gives the dynamics of only one degree of freedom, namely the vertical position of the tip $z(t)$.  It is therefore well-motivated to formulate reduced models which approximate the tip-sample collision dynamics.  

The basic assumption of nearly all models in AFM is that the tip-sample interaction force can be expressed as a function of two dynamic variables,
\begin{equation}
    F_\text{TS}(t) = f \left(z(t),\dot{z}(t) \right).
    \label{eq:f(z,zdot)}
\end{equation}
We argue that this basic assumption is incorrect for soft materials as it neglects the fact that the sample itself also has dynamics, which clearly influence the tip-sample interaction.  Our approach is to reduce the complicated three-dimensional dynamics of the sample to a simplified one-dimensional dynamics in terms of a single degree of freedom $z_s(t)$, a generalized or spatially-averaged vertical position of the surface in the laboratory frame [see Fig.~\ref{fig:fts}b)].  If we neglect the inertia of the sample contained in the very small interaction volume, our interaction force depends on this one additional 'hidden' dynamic variable
\begin{equation}
    F_\text{TS}(t) = f \left(z(t),\dot{z}(t) , z_s(t) \right).
\end{equation}
This type of model was introduced by Cantrell and Cantrell to account for externally forced oscillations of the sample in the context of ultrasonic AFM.~\cite{Cantrells_2013}  

Below we describe a simplified version of the moving surface model. A more complicated variation was presented and compared to experiments on soft materials in our previous publication~\cite{Haviland:SoftViscoElastic:2015}.  The simpler model has analytic solutions in special cases which give physical intuition.  We simulate the simplified model and introduce numerical optimization to find the model parameters that best fit experimental data.  We argue that the traditional approach to quantitative AFM, where model parameters are bulk material properties such as elastic modulus and viscosity, is not appropriate for soft materials because a physically correct and complete model would involve far too many free parameters and uncontrolled assumptions.  Rather, we demonstrate that a dimensionless ratio of our model parameters, called the impact-rheology factor $R$, gives a  physically relevant and useful quantity for nano-scale mechanical characterization of viscoelastic surfaces.  Our measurements show that the impact-rheology factor is independent of oscillation frequency and details of the tip geometry.

\section{The moving surface model}

All coordinates of the moving surface model are defined in Fig.~\ref{fig:fts}b.  The essential difference to the traditional approach (Fig.~\ref{fig:fts}a) is that we introduce the instantaneous position of the surface $z_s(t)$ as an independent dynamic variable and we express the interaction force as a function of the tip-surface separation $s=z-z_s$.  Unlike traditional models, we do not require that the tip and surface move together ({\em i.e.} $z=z_s$) when they are in contact.

One could imagine many different nonlinear models for the interaction force, but we make a sweeping simplification by linearizing the interaction when in contact.  Figure~\ref{fig:fts}c shows a piece-wise linear (PWL) force model described by the following equations, 
\begin{equation}
 F_\text{TS}(s,\dot{s}) =
  \begin{cases}
    0 & \text{if  } s > 0\\
   -F_\text{ad} - k_\text{v} s - \eta_\text{v} \dot{s}  & \text{if  } s \leq 0
  \end{cases},
  \label{eq:fts}
\end{equation}
The interaction force is zero when the tip is out of contact, $s>0$.  When in contact, $s\leq0$, the adhesion force $F_\text{ad}$ turns on, correspond to a lowering of the total interfacial energy.  Adhesion is counteracted by a repulsive force linear in the penetration $-s$, with force constant $k_\mathrm{v}$.  We also include the possibility of a viscous force, linear in $\dot{s}=\dot{d}-\dot{d}_\mathrm{s}$ ($h$ and $z_0$ are constant) associated with material flow as the tip penetrates the sample.

The model preserves one essential feature of the interaction which is well known in AFM -- the large force gradient $\frac{\mathrm{d}}{\mathrm{d}z}F_\text{TS}(s, \dot{s})$ localized near the point of contact $s=0$.  This rapid change of force is responsible for the jump-to-contact and pull-off hysteresis seen in nearly all quasi-static force curves, or measurements of $d(h)$ for small $\dot{h}$, traditionally analyzed in quasi-static AFM.  In dynamic AFM the amplitude of oscillation is typically much larger than the range of this localized interaction.  We may therefore approximate this region of large interaction gradient as an adhesion force which instantly turns on and off when crossing the point of contact $s=0$.

The interaction force couples the dynamics of the cantilever's flexural eigenmode to the dynamics of the viscoelastic surface, described by the following set of equations
\begin{subequations}
\begin{align}
m \ddot{d} + \eta \dot{d} + k d  =&  F_\text{TS}(s, \dot{s})  + F_\text{drive}(t) \label{eq:cant} \\
\eta_\text{s} \dot{d_\text{s}} + k_\text{s} d_\text{s}  =&  -F_\text{TS}(s, \dot{s}) \label{eq:surf}.
\end{align}
\end{subequations}
Here $d_\text{s} = z_\text{s} - z_0$ is the deflection of the surface from its equilibrium position $z_0$ and $F_\text{drive}(t)$ is the drive force.

The free motion of the tip is given by eq.~\eqref{eq:cant} with $F_\text{TS}=0$, describing a driven damped harmonic oscillator.  This model is valid in a narrow frequency band around a high quality factor resonance of the cantilever.  The three mode parameters: stiffness $k$, resonant frequency $\omega_0 = 2\pi f_0 = \sqrt{k/m}$ and dimensionless quality factor $Q= \omega_0 \tau =\sqrt{\eta^2/mk}$, are independently determined using a calibration procedure (see Methods). 

The free motion of the surface is described by eq.~\eqref{eq:surf} with $F_\text{TS}=0$.  Note that eq.~\eqref{eq:surf} does not have an inertial term corresponding to force arising from acceleration of the sample mass.  Neglecting sample mass is valid when deformation occurs only in a local volume around the tip, however inertial forces may arise if the cantilever excites standing surface waves~\cite{Haviland:SoftViscoElastic:2015}.   Thus the model effectively puts the center-of-mass at the equilibrium position of the tip, which we treat as fixed in the laboratory frame.  We therefore neglect the very small amplitude base motion needed to inertially actuate the high Q resonance~\cite{Platz:RoleNLDynamics:2012}, as well as any forces arising from rapid changes of the probe height due to overly active surface-tracking feedback.    

\begin{figure}[ht]
\begin{center}
\includegraphics[width=7cm]{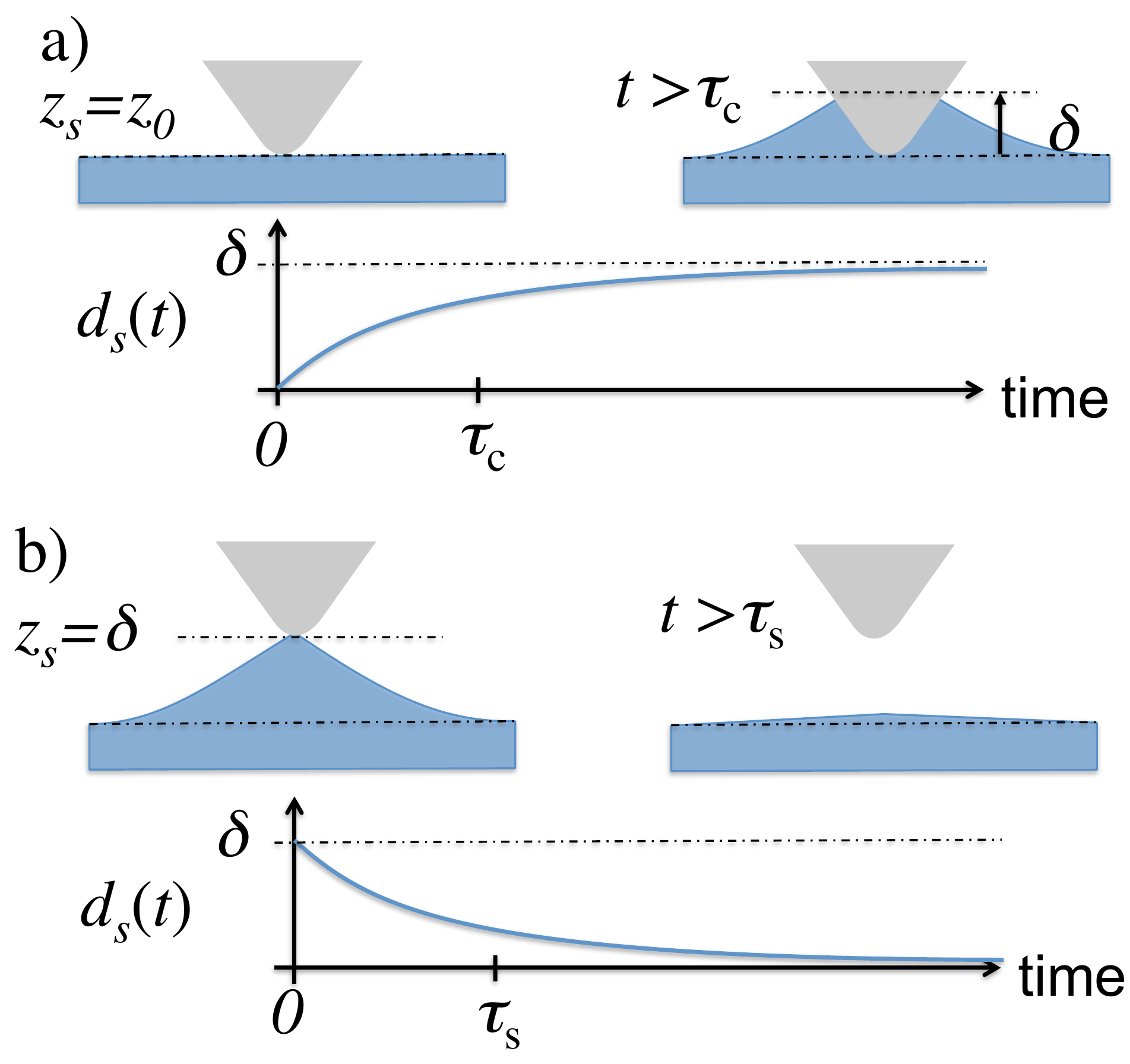}
\caption{{\bf Contact formation and free relaxation}.  (\textbf{a}) At $t=0$ a tip, rigidly fixed in the laboratory frame, meets the sample surface.  The adhesion force turns on and the surface lifts by an amount $\delta$, forming the contact in a characteristic time $\tau_\text{c}$.  (\textbf{b}) At $t=0$ the tip separates from the lifted surface.   The surface relaxes to its equilibrium position in the characteristic time $\tau_\text{s}$. }
\label{fig:touch}
\end{center}
\end{figure}

We can develop intuition for the sample dynamics by considering what happens when the tip is held rigidly fixed in the laboratory frame  (\textit{i.e.} not connected to a flexible cantilever) at the height of the unperturbed surface.  As the tip just touches the surface from above [see Fig.~\ref{fig:touch}(a)], the case $s=0$ in eq.~\eqref{eq:fts} together with the condition $z = z_0$, or equivalently $s = -d_s $ in eq.~\eqref{eq:surf}, gives
\begin{equation}
(\eta_\text{s} + \eta_\text{v})\dot{d_\text{s}} + (k_\text{s} + k_\text{v})d_\text{s} = F_\text{ad} .
\end{equation}
Solving this equation we find that upon contact, the adhesion force lifts the surface
\begin{equation}
d_\text{s}(t) = \delta ( 1 - e^{-\frac{t}{\tau_\text{c}}} )
\end{equation}
forming a meniscus with asymptotic height,
\begin{equation}
\delta =  F_\text{ad}/(k_\text{v} + k_\text{s}) .
\label{eq:delta}
\end{equation}
in a characteristic contact formation time,
\begin{equation}
\tau_\text{c} = \frac{\eta_\text{v} + \eta_\text{s}}{k_\text{v} + k_\text{s}}. 
\end{equation}
Similarly, when the tip just separates from the lifted surface, eq.~\eqref{eq:fts} for the case $s>0$ and eq.~\eqref{eq:surf} give,
\begin{equation}
\eta_\text{s} \dot{d_\text{s}} + k_\text{s} d_\text{s} = 0 .
\end{equation}
describing free relaxation of the surface to its equilibrium position in a characteristic time
\begin{equation}
\tau_\text{s} =  \eta_\text{s} /  k_\text{s}. 
\end{equation}
The contact formation and free relaxation dynamics are depicted in Fig.~\ref{fig:touch}.  We may also define the time constant 
\begin{equation}
   \tau_v = \frac{\eta_v}{k_v},
\end{equation}
associated with tip penetration into the sample.

The model behaves as we would intuitively expect for an interaction consisting of two opposing forces: attractive surface force resulting from minimization of interfacial energy, and repulsive volumetric force resulting from compressive stress in the bulk.  The stiffness parameters $k_\text{s}$ and $k_\text{v}$, include both surface and bulk forces with the relative contribution depending on the size of the deformation in relation to the elastocapillary length.  Nevertheless, a liquid-like interaction is described by $K \equiv k_\text{s} / k_\text{v} \gg 1$, in which case the meniscus lifts to a maximum height,
\begin{equation}
\delta \simeq \delta_0 \equiv F_\text{ad}/k_\text{s}
\label{eq:delta0}
\end{equation}
A solid-like interaction is described by the opposite limit $K \ll 1 $, in which case $\delta \ll \delta_0$ as capillary adhesion is counteracted by compressive stress in the contact volume.  

The time constants $\tau_\text{c}$, $\tau_\text{s}$ and $\tau_\text{v}$ represent ratios of viscous to elastic force constants of the model.  The dynamics of both the tip and the sample depend on these characteristic times scales and their relation to the time spent in and out of contact during a single oscillation cycle, where the latter is set by the frequency of cantilever oscillation, the amplitude of motion and working distance to the surface $w=h-z_0$.  When the contact formation time $\tau_\text{c}$  is larger than the time spent in contact, the surface can not fully deform to achieve contact equilibrium.  When the free relaxation time $\tau_\text{s}$ is large compared to time spent out of contact, the surface can not relax to its equilibrium position before the next tap of the tip.  Repeated taps result in a steady-state dynamics characterized by a time-average up-lifted or indented position of the surface. 

In the following section we demonstrate this behavior of the model and correlate it with data from experiments.  We show that this simple model explains the experimental data remarkably well for a variety of different soft samples.  Fitting the model to experimental data we extract the parameters which characterize the material and its surface.  We then comment on how these parameters may be useful for a general, quantitative characterization of viscoelastic samples with AFM.

\section{Comparison with experiment}

\begin{figure}[htb]
 \center
\includegraphics[width=8 cm]{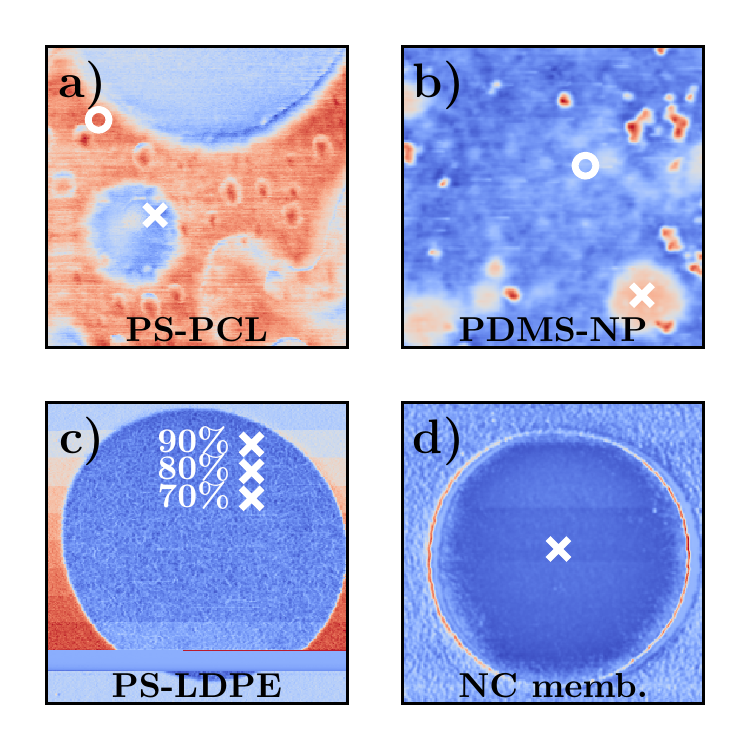}
\caption{\textbf{Phase images at the first drive frequency of ImAFM.}  (\textbf{a}) Scan size 0.8~$\mu$m.  Domains of Polycaprolactone (PCL, blue) in Polystyrene (PS, red).  Pixels marked with \textbf{o} and $\times$ are analyzed in Fig.~\ref{fig:PCL_FIFQ}.  (\textbf{b})  Scan size 1~$\mu$m.  Polydimethylsiloxane (PDMS, blue) with clusters of hydrophobic silica nanoparticles (NP, red). Pixels marked with \textbf{o} and $\times$ are analyzed in Fig.~\ref{fig:pdms_nanoparticles_fifq}. (\textbf{c})  Scan size 1.5~$\mu$m.  A setpoint study on a domain of low density polyethylene (LDPE, blue) in PS.  Pixels marked with $\times$ are analyzed in Figs.~\ref{fig:Tap525_soft}, S1 and S2, at the setpoint values shown.  (\textbf{d})   Scan size 3~$\mu$m.  A suspended membrane of Au nanoparticles bound together in a monolayer by organic ligands.  A pixel at the center of the membrane marked with $\times$ is analyzed in Fig.~\ref{fig:nanocomp_mem_fifq}.}
\label{fig:samples}
\end{figure}

We have validated the model by fitting it to data collected on several different samples.  Figure~\ref{fig:samples} shows scanned images of four samples discussed below.   To display the results of both experiment and theory, we show dynamic force quadrature curves using a technique called Intermodulation AFM (ImAFM)~\cite{Platz:ImAFM:08}.  Force quadratures have been introduced in previous publications~\cite{Platz:InteractionImaging:2013,Haviland:COCISreview:2017} but they are rather unfamiliar to the majority of AFM scientists so we briefly describe them here.    Force quadrature curves do not display the instantaneous force on the tip as a function of tip position $ F_\text{TS}(z)$ ({\em i.e.} 'force curves').  Rather, force quadratures represent integrals of force over a single oscillation cycle, making them the natural force curves of dynamic AFM.  

The reader familiar with dynamic mechanical analysis (DMA) will recognize a similarity between force quadratures and typical DMA experiment.
In DMA, the strain in a bulk material is measured while a sinusoidal stress of fixed amplitude is applied.
The in-phase and quadrature components of the strain are then monitored as a function of the frequency of excitation (or the temperature of the sample) to characterize the complex modulus $E^* = E^\prime + \mathrm{i}E^{\prime\prime}$ of the material under test.
To determine the AFM force quadratures we instead monitor the in-phase and quadrature response of the cantilever, subject to a sinusoidal excitation at fixed frequency but with slowly varying amplitude.  
This allows us to study how the nonlinear features associated with the impact of the tip on the viscoelastic sample change with amplitude.  To explain these features we must look beyond the bulk concepts of storage and loss moduli to also include surface forces and adhesion.


The high quality factor of the cantilever oscillations means that the stored energy in the cantilever oscillation is much larger than the tip-sample interaction potential or energy lost to dissipation in the cycle.  The tip motion is therefore well-described by harmonic oscillation at a fixed 'carrier' frequency $\bar{\omega}   \simeq \omega_0$.  The second drive tone of ImAFM gives rise to a slowly-modulated amplitude and phase of the carrier.  From the intermodulation spectrum of these two drive tones, which is concentrated near resonance, we extract the Fourier coefficients of the tip motion and tip-sample force, at this carrier frequency.  Rotating the force coefficients by the motion phase, we project out the Fourier coefficients of force which are in phase with the motion $F_I$, and that which are quadrature to the motion $F_Q$~\cite{Platz:InteractionImaging:2013,Haviland:COCISreview:2017}.  Both are presented as functions of oscillation amplitude $A$.
\begin{subequations}
\begin{align}
z(t) &= h + A \cos \bar{\omega} t    \label{eq:d(t)}  \\
F_I(A) &= \frac{1}{T} \int_0^T F_{TS}(t)  \cos (\bar{\omega} t) dt    \label{eq:FI}  \\
F_Q(A) &= \frac{1}{T} \int_0^T F_{TS}(t)  \sin (\bar{\omega} t) dt   \label{eq:FQ}  
\end{align}
\end{subequations}
 $F_I$ represents a conservative force (\textit{i.e.} energy recovered in the oscillation cycle), and $F_Q$ represents a dissipative force (\textit{i.e.} energy lost in the cycle).   

When the tip is tapping on the sample, large Fourier components of the tip-sample force also exist at frequencies far above resonance but well below the detection noise floor. Thus, we can not determine the instantaneous force at any given time in the cycle.  However,  using the multifrequency lockin technique embodied in ImAFM, we can determine the integrated force over single cycles with very good signal-to-noise ratio.  The two curves $F_I(A)$ and $F_Q(A)$  are found by direct transformation of the intermodulation spectrum, requiring only the calibrated linear response function of the cantilever eigenmode.  No assumptions are made regarding the specific nature of the tip-sample interaction. The transformation is quite efficient computationally, allowing for immediate examination of the force quadrature curves at individual pixels, while scanning.

\begin{figure*}[htb]
\includegraphics[width=\textwidth]{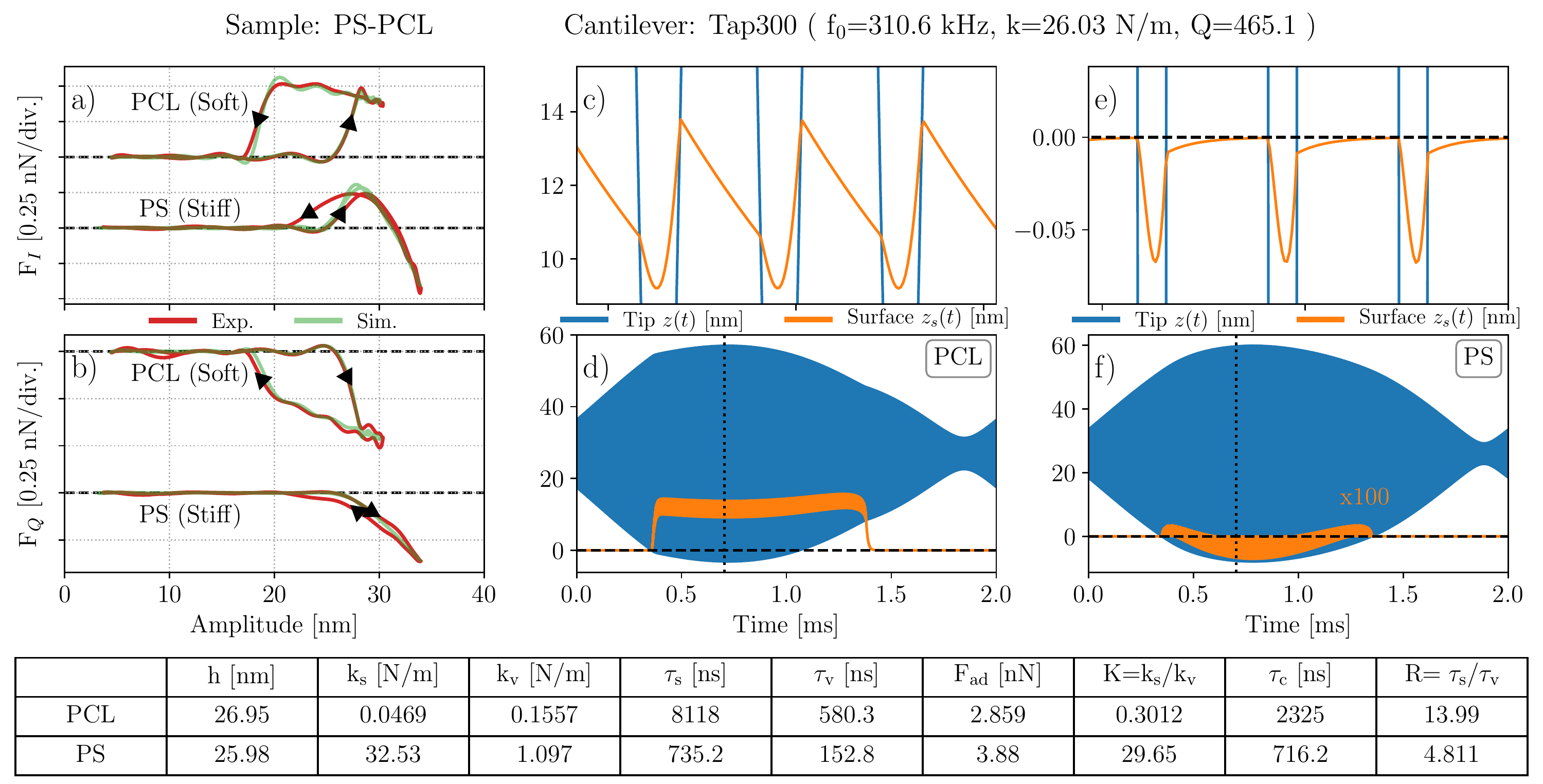}
\caption{\textbf{Polystyrene/polycaprolactone blend.} (\textbf{a}) The conservative force quadrature $F_I(A)$ and (\textbf{b}) the dissipative quadrature $F_Q(A)$.  The curves are offset vertically for clarity with dashed lines corresponding to zero force.  The simulated motion of the tip $z(t)$ (blue) and surface $z_s(t)$ (orange) are shown for both PCL (\textbf{c} - \textbf{d}) and PS (\textbf{e} - \textbf{f}).  The zooms (\textbf{c}) and (\textbf{e}) showing details of the surface motion around the time marked by the vertical dotted line in (\textbf{d}) and (\textbf{f}) respectively.  The surface motion of PS in (\textbf{f}) is magnified by a factor of 100.  Note that the average surface position for PCL is $\sim 12$~nm above its rest position, while PS deviates only $\sim 0.05$~nm from its equilibrium position.  The best-fit parameters used in simulation are given in the table.}
\label{fig:PCL_FIFQ}
\end{figure*}

Figure~\ref{fig:PCL_FIFQ} shows force quadrature curves taken at two different points on a blend of polystyrene (PS) and polycaprolactone (PCL), shown in Fig.~\ref{fig:samples}(a).  The force quadratures on the softer PCL (nominal bulk $E\sim$300~MPa) show larger magnitude of the dissipative force $F_Q$ and a conservative force which is dominantly attractive, $F_I>0$, even at the highest amplitude.  To fit the theory to the experimental data, we simulate the model dynamics by numerical integration of the equations of motion eq.~\eqref{eq:cant} and eq.~\eqref{eq:surf}, adjusting the parameters of the model to find a best fit.   Details of this procedure are given in the Methods section.  

On the soft PCL we find an excellent fit. The simulation accurately reproduces the magnitude, complex shape and hysteresis of both $F_I(A)$ and $F_Q(A)$. The simulated surface motion shows that the adhesive force causes the surface to lift by a time-averaged distance of about 15~nm above its equilibrium position  (fig.~\ref{fig:PCL_FIFQ}d).  The zoom of individual oscillation cycles (fig.~\ref{fig:PCL_FIFQ}c) shows that upon impact the tip penetrates as much as $\sim$20~nm  below the lifted surface.  After separation on the upward trajectory of the tip, the surface does not have time to fully relax to its equilibrium position before the next impact.  This slow relaxation of the surface gives rise to the hysteresis observed in the $F_I(A)$ and $F_Q(A)$ curves:  the amplitude at which the surface is initially lifted up [sharp up-turn in $F_I(A)$], is larger than the amplitude where the oscillating tip fully releases the surface [sharp down-turn in $F_I(A)$]. This hysteresis is not easily explained by a model having the form of Eq.~\eqref{eq:f(z,zdot)}, giving confidence that the basic physical picture embodied in the moving surface model is correct.

We compare the behavior on the softer PCL with that on the stiffer PS domain (expected $E\sim$3 GPa).  On PS we find that the conservative force quadrature $F_I$ becomes dominantly repulsive  ($F_I<0$) at large amplitude, and we see a smaller magnitude of the dissipative force $F_Q$.  Simulations show that the tip penetrates as much as 10~nm below the surface, but the surface lifts only slightly and quickly relaxes to its equilibrium position before the next impact.    For increasing $A$ the simulation captures the shape and magnitude of both $F_I(A)$ and $F_Q(A)$.  However for decreasing $A$ we find that the simulation does not reproduce the hysteresis observed in the experiment.  A different interaction function $F_\text{TS}(s)$ could improve the quality of the fit, but at the expense of introducing additional model parameters.  In our experience, the simplified model presented herein often does not capture hysteresis observed together with $F_I<0$, but it nearly always captures hysteresis when $F_I>0$.  


\begin{figure*}[!htb]
\includegraphics[width=\textwidth]{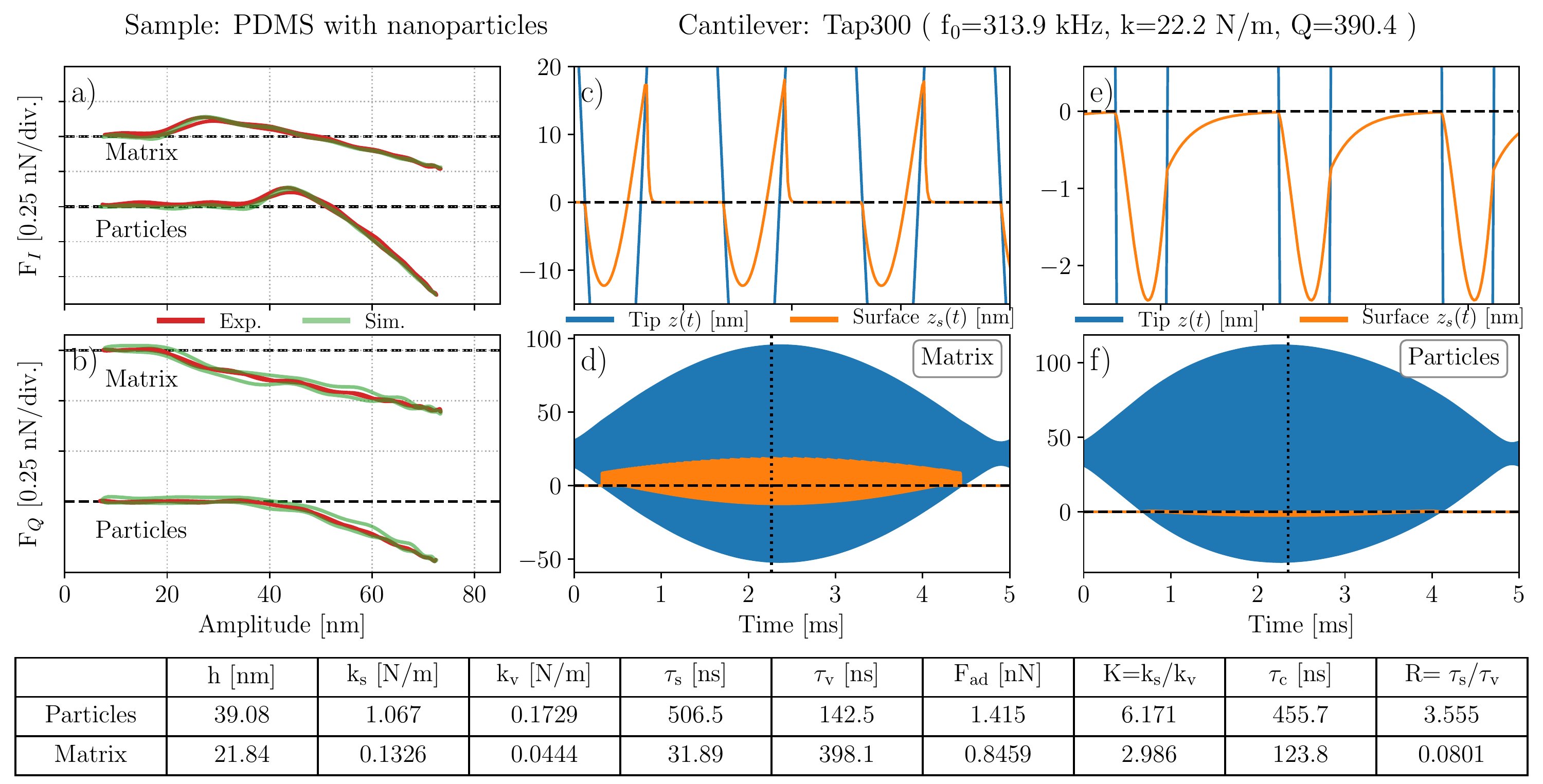}
\caption{\textbf{PDMS with hydrophobic silica nanoparticles}. The details of this sample are described in a previous publication~\cite{Huang:PDMSnanoparticles:2017}.  The curves labeled Matrix were measured on the softer PDMS, where the simulation shows large amplitude surface motion and deep tip penetration.  The curves labeled Particles were measured on a subsurface aggregation of nanoparticles, where the simulation reveals very little surface motion and lifting, yet equally deep tip penetration.   \textbf{(a)} The conservative force quadrature $F_I(A)$ and \textbf{(b)} the dissipative quadrature $F_Q(A)$.  The curves are offset vertically for clarity with dashed lines corresponding to zero force.  The simulated motion of the tip $z(t)$ (blue) and surface $z_\text{s}(t)$ (orange) are shown for both the PDMS matrix (\textbf{c}) and (\textbf{d}) and a cluster of particles (\textbf{e}) and (\textbf{f}).  The zooms (\textbf{c}) and (\textbf{e}) showing details of the surface motion around the time marked by the vertical dotted line in (\textbf{d}) and (\textbf{f}) respectively.   The best-fit parameters used in the simulation are given in the table.}
\label{fig:pdms_nanoparticles_fifq}
\end{figure*}

Figure~\ref{fig:pdms_nanoparticles_fifq} shows an additional example of model fitting at two locations on a sample consisting of PDMS mixed with hydrophobic silica nanoparticles, shown in Fig.~\ref{fig:samples}b.  This soft sample has faster free relaxation (smaller $\tau_s$) such that no hysteresis is seen in the force quadratures.  The model reveals large amplitude surface motion when tapping on the PDMS matrix.  When tapping on a region with dispersed nanoparticles, the model shows a much slower surface and very low amplitude surface motion, but equally deep penetration.   

We also investigate how the fitted model parameters depend on a change of tip and cantilever oscillation frequency.  Using three standard cantilevers with calibrated parameters given in Table~\ref{tab:LDPE_harmonix}, we study a well-known soft material consisting of micron size domains of LDPE (nominal bulk $E\sim$100~MPa) in a matrix of PS~\cite{BrukerHarmoniX}, as shown in Fig.~\ref{fig:samples}c .   The cantilever resonant frequencies and mode stiffnesses span one order of magnitude. In order to make a reasonable comparison we adjust the excitation to keep the amplitude of free motion $A_\text{free}$ such that the stored energy in the free oscillation $E_\text{free}=\frac{1}{2}kA_\text{free}^2$  is approximately the same in each measurement.   Table~\ref{tab:LDPE_harmonix} also shows the best-fit parameters of the moving surface model. 

\begin{figure}
\centering
\includegraphics[width=8 cm]{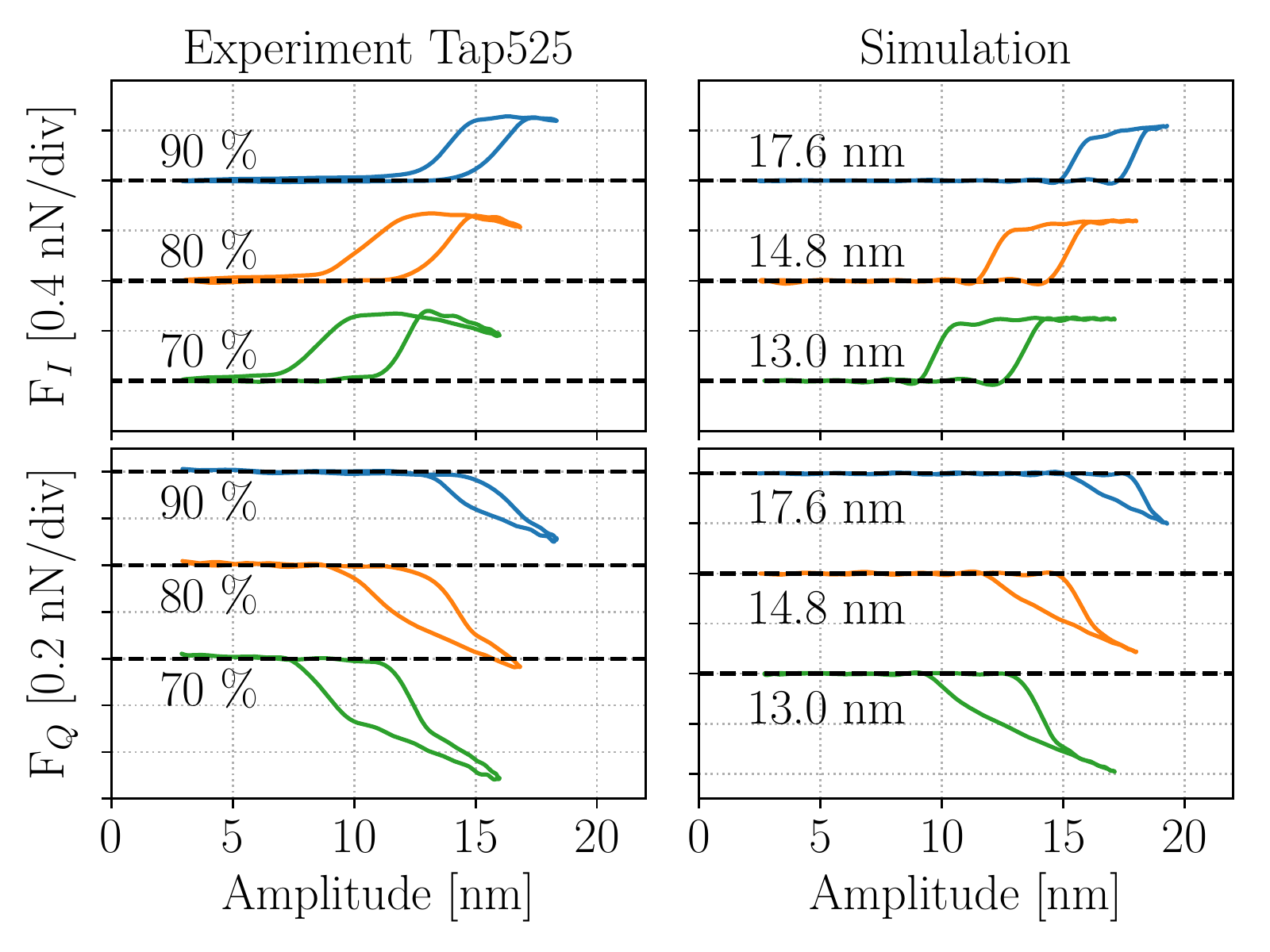}
\caption{{\bf Low-density polyethelene (LDPE) with Tap525 cantilever}. Experimental (left) and simulated (right) force quadrature curves, each offset vertically for clarity with dashed lines corresponding to zero force.  The experimental curves result from analysis of data at pixels marked with an $\times$ of the corresponding color in Fig.~\ref{fig:samples}(c).  The working distance is changed in the experiment by adjusting the amplitude-feedback setpoint to the value given in left panels ({\%} of free amplitude).  For simulation the working distance $w$ shown in the right panels is found by numerical optimization with all other parameters of the model fixed at the values given in Table~\ref{tab:LDPE_harmonix}. }
\label{fig:Tap525_soft}
\end{figure}

For each cantilever we also study how the force quadrature curves change as we vary the working distance, $w=h-z_0$, or static probe height above the relaxed surface.  We control $w$  by simply changing the scanning feedback set-point at regular intervals along the slow scan axis [see Fig.~\ref{fig:samples}(c)].  Figure~\ref{fig:Tap525_soft} shows an example of the measured and simulated force quadratures for the Tap525 cantilever at three different probe heights.  All simulation curves shown in the figure use the same parameters given in Table~\ref{tab:LDPE_harmonix}, changing only the working distance $w$ which is given in Fig.~\ref{fig:Tap525_soft}.  Equivalent plots for the NSC15 and AC55 cantilevers are shown in the supplemental material Figs.~S1 and S2 respectively.  We see that the model reproduces the shape and offset of the force quadrature curves rather well.

\begin{table}
\centering
\begin{tabular}{| c | c | c | c |}
    \hline
    Cantilever          & \textbf{NSC15}        & \textbf{Tap525}       & \textbf{AC55} \\ \hline
    $f_0$ [kHz]         & 230                   & 489                   & 2120 \\ \hline
    $Q$                 & 397                   & 547                   & 735 \\ \hline
    $k$ [N/m]           & 18.9                  & 88.6                  & 161 \\ \hline
    $A_\mathrm{free}$ [nm] & 20                 & 11                    & 6.3 \\ \hline
    \hline
    $F_\text{ad}$ [nN]  & 0.955                 & 1.8                   & 0.818 \\ \hline
    $k_s$ [N/m]         & 0.17                  & 0.858                 & 0.227 \\ \hline 
    $\eta_s$ [kg/s]     & 3.01$\cdot10^{-6}$    & 14.9$\cdot10^{-6}$    & 2.75$\cdot10^{-6}$ \\ \hline 
    $k_v$ [N/m]         & 0.04                  & 0.0238                & 0.014 \\ \hline 
    $\eta_v$ [kg/s]     & 8.24 $\cdot10^{-8}$   & 4.74 $\cdot10^{-8}$   & 1.92 $\cdot10^{-8}$ \\ \hline
    \hline
    $\tau_s$ [$\mu$s]   & 17.68                 & 17.31                 & 12.11 \\ \hline
    $\tau_v$ [$\mu$s]   & 2.06                  & 1.99                  & 1.37 \\ \hline
    $\tau_c$ [$\mu$s]   & 14.7                  & 17.1                  & 11.5 \\ \hline 
    $K$                 & 4.25                  & 3.6                   & 16.21 \\ \hline   
    $R$                 & 8.61                  & 8.69                  & 8.86 \\ \hline 
\end{tabular}
\caption{{\bf Parameters for LDPE with three different cantilevers}.  Calibrated cantilever parameters (top box) and fitted model parameters (middle box).  Ratios of the model parameters (bottom box) are: the relaxation times $\tau_s=\eta_s/k_s$ and $\tau_v=\eta_v/k_v$, contact formation time $\tau_\text{c}=(\eta_\text{s} + \eta_\text{v}) / (k_\text{s} + k_\text{v}) $, stiffness ratio $K=k_s/k_v$  and impact-rheology factor $R=\tau_s/\tau_v$.  Although the stiffness and damping parameters of the model show considerable variation between probes with different frequency and tip geometry, the time constants show much less variation.  The impact-rheology factor $R$ shows little variation between probes.  }
\label{tab:LDPE_harmonix}   
\end{table}{}

\begin{figure*}[htb]
\includegraphics[width=\textwidth]{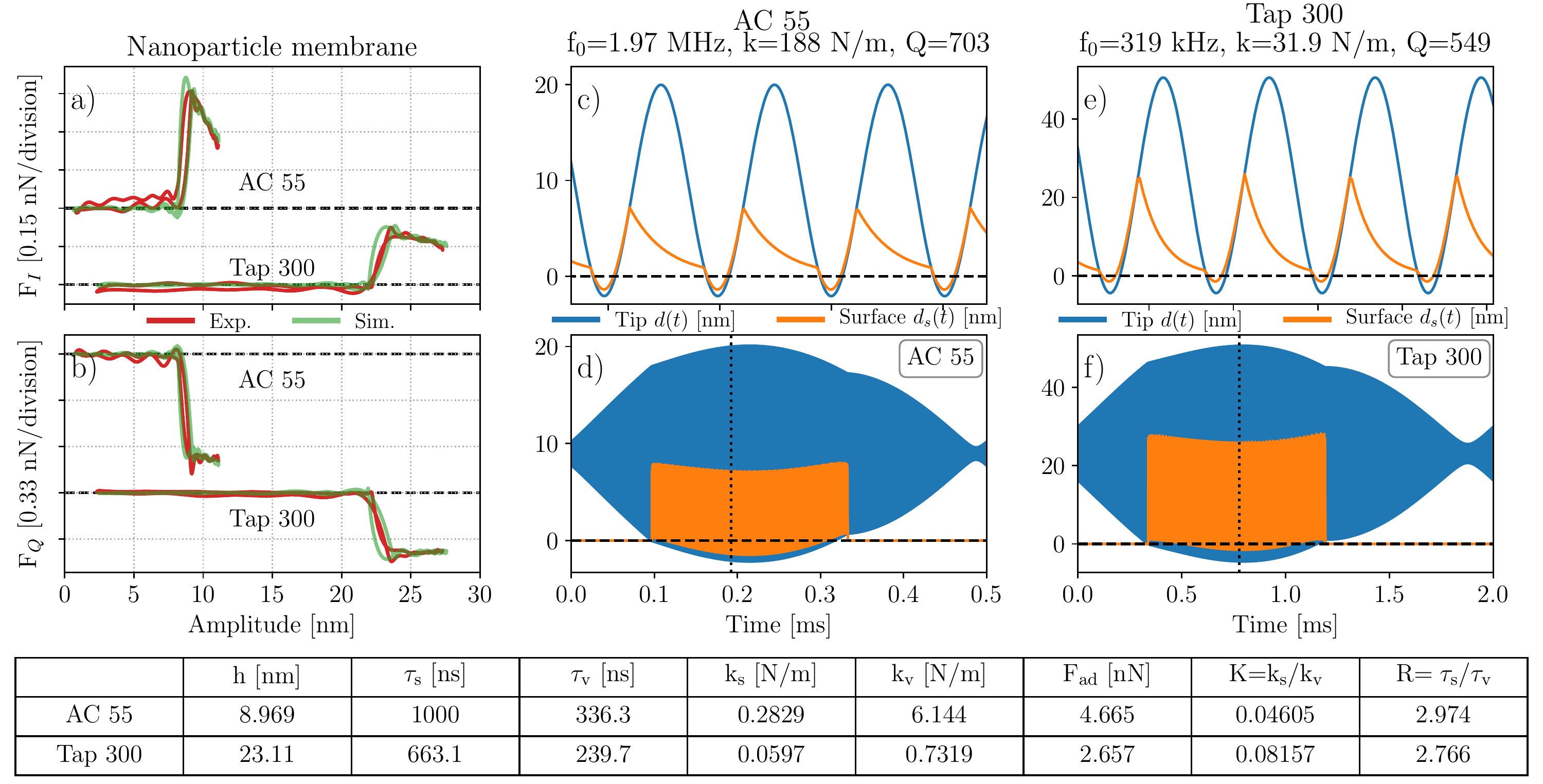}
\caption{\textbf{Nanoparticle membrane with Tap300 and AC55}.   Measurements are made in the middle of a membrane with diameter 2.5$\mu$m.  We show results from two different membranes from the same batch, measured with two different cantilevers.   (\textbf{a}) The conservative force quadrature $F_I(A)$ and (\textbf{b}) the dissipative quadrature $F_Q(A)$.  The curves are offset vertically for clarity with dashed lines corresponding to zero force.  (\textbf{c}) and (\textbf{d}) show the simulated motion of the tip $z(t)$ (blue) and surface $z_\text{s}(t)$ (orange) for cantilever AC55, and (\textbf{e}) and (\textbf{f}) for cantilever Tap300.  The zooms (\textbf{c}) and (\textbf{e}) show details of the surface motion around the time marked by the vertical dotted line in (\textbf{d}) and (\textbf{f}) respectively.   The best-fit parameters used in the simulation are given in the table. }
\label{fig:nanocomp_mem_fifq}
\end{figure*}

To further test the model we also analyze a nanoparticle membrane sample shown in Fig.~\ref{fig:samples}(d). This sample consists of a monolayer of Au nanoparticles, bound together by organic ligands.  The membrane is suspended over a circular hole 2.5$\mu$m in diameter, forming a ultra-thin drum head~\cite{Jaeger:NPMembraneSmall:2010,JagerSader:Drumheads:2013}.  Previous studies have shown large apparent stiffness, corresponding to a large effective bulk modulus of the membrane material~\cite{Wang:NPmembrane:2017}. Interpretation of the model parameter $k_\text{v}$ in terms of bulk compression is not obvious with these samples that have thickness $\simeq 7$nm.  However, adhesion gives rise to very large curvature of upper and lower surfaces of the membrane as it covers the tip with very small radius.  The Laplace pressure change across both of these interfaces adds, giving the force described by $k_\text{v}$ which acts in opposition to adhesion.  The stiffness constant $k_\text{s}$ then describes elastic tension in the membrane as the surface deviates from equilibrium.  It is therefore not surprising that our simple model fits the data quite well.  Figure~\ref{fig:nanocomp_mem_fifq} shows a comparison of experiment and simulation of the model for two different cantilevers. The qualitative shape of the force quadratures and the simulated dynamics of the surface are the same for both cantilevers, despite the difference in resonance frequency of AC55 ($\sim$~2~MHz) and Tap300 ($\sim$~300~kHz).

\section{Discussion}

An obvious criticism of the model presented herein is that it is {\it ad hoc}, or not based on first principles.   Nonlinear forces arising from sample deformation, the analysis of which forms the basis of traditional AFM nanomechanics, are avoided by linear approximation.  The model keeps only one nonlinearity to capture the sudden impact and release of the tip and sample.  Reduction of the sample dynamics to one effective degree of freedom $d_s(t)$ is a sweeping simplification which neglects the fact that the surface deformation has a transverse profile.   

We emphasize again that our intent is to describe the measured dynamics of the cantilever with a minimum number of free parameters and additional degrees of freedom.  A physically more realistic short-range interaction force which smoothly interpolates between the two linear regions could be constructed, but at the expense of additional parameters which are not necessary to explain the data.  We consider this to be the minimal model necessary for explaining dynamic AFM on viscoelastic materials, and in our experience it is often sufficient.  
    
Starting from first principles, even very idealized models would involve many parameters:  tip radius and side-wall angle - assuming an axial-symmetric tip; elastic modulus, Poisson ratio and viscosity - assuming homogeneous half-space; surface energies and local curvature of the unperturbed surface.  All parameters not independently determined should be treated as free when fitting AFM data.  The fit procedure here uses the multifrequency ImAFM data at some 40 frequencies near resonance.  While the quadrature data (80 values) do have different weight (they are measured with different signal-to-noise ratio) each represents an independent observable with information about the tip-sample interaction.  In stark contrast to other quantitative methods in AFM, the large number of observables in relation to the number of free model parameters makes physical interpretation of the fitted parameter values meaningful.  

Examining Table~\ref{tab:LDPE_harmonix}, we recall that the cantilever parameters $f_0$, $Q$ and $k$ are independently calibrated (see Methods),  and the working distance $w$ can be determined by inspection (amplitude at the initial sharp rise in $F_I$).  Thus the 5 parameters $F_\text{ad}$, $k_\text{s}$, $\eta_\text{s}$, $k_\text{v}$ and $\eta_\text{v}$, are free parameters.  We expect that all of these will depend on the detailed shape of the tip.  It is therefore not so surprising that the table shows considerable variation of these parameters between the different probes.  We expect that the relaxation time $\tau_\text{s}= \frac{\eta_\text{s}}{k_\text{s}}$ is independent of tip shape as it involves the free relaxation of the sample.  Furthermore $\tau_\text{v}= \frac{\eta_\text{v}}{k_\text{v}}$ might, to first order, be sensitive to tip shape as a blunt tip would result in both larger viscous and larger elastic force, in comparison with a sharper tip.  Indeed, we do observe little variation in the values of $\tau_\text{s(v)}$ for the two cantilevers with lower frequency, but there is a reduction of both $\tau_\text{s}$ and $\tau_\text{v}$ for the higher frequency cantilever.   
 
 It is interesting to note in Table~\ref{tab:LDPE_harmonix} and in the table of Fig.~\ref{fig:nanocomp_mem_fifq}, that the ratio of time constants
 \begin{equation}
 R = \frac{ \tau_\text{s} }{ \tau_\text{v} } = \frac{ \eta_\text{s}  k_\text{v}  }{ k_\text{s}   \eta_\text{v} } 
 \end{equation}
is apparently independent of the AFM probe used in the experiment. We call $R$ the impact-rheology factor~\cite{RheologyImpactFactor}, a dimensionless number formed from all four force constants of our PWL viscoelastic model.  Our measurements suggests that this factor may be a good quantity for physical characterization of surface viscoelasticity.  

In order to quantify measurement uncertainty in $R$,  one should study how various sources of noise in the experiment propagate to the model parameters in the fitting procedure.  Further experimental studies and more detailed model analysis are required to study uncertainties, and verify if the impact-rheology factor is indeed a reliable measurement of an intrinsic property of the sample interface.   

While this moving surface model is minimal with regard to the number of parameters, it nevertheless has very complicated dynamics.  General statements about its behavior in different parameter regimes are therefore difficult to formulate.  The impact-rheology factor represents the relative strength of two viscoelastic processes:  large $R$ means that the sample's free relaxation is slow in comparison to the time needed for the penetrating tip to locally deform the sample.  Conversely, small $R$ means that tip penetration is slow in comparison with free relaxation.  However, the magnitude of $R$ alone is insufficient to predict the shape of the force quadrature curves, which also depend on the adhesion force, working distance, and frequency of oscillation.   Hysteresis is associated with $\omega_0 \tau_\text{s}\gg 1$, large surface lifting with small $k_\text{s}$ or large $\delta_0$, and deep penetration with small $k_\text{v}$.

\section{Conclusions}

We presented a minimal model for dynamic AFM on viscoelastic materials, accounting for both surface and bulk forces. The model is simple in that it is linear where possible, but its dynamics is complex due to the nonlinearity describing the sudden impact and release of the oscillating tip and surface. Unlike traditional nanomechanical analysis of AFM, our model takes into account the viscoelastic dynamics of both the penetrating tip and free sample.  The sample dynamics was shown to be quite significant on soft materials, where simulations revealed large amplitude surface motion.  We validated the model by showing excellent agreement with experimental data on a variety of samples.  By fitting the model to the data we extracted viscous and elastic force constants for the surface and bulk.  Our analysis indicated that the impact-rheology factor, formed from the dimensionless ratio of these constants, is independent of tip shape and cantilever resonance frequency.  This simple model describes numerous measured force quadrature curves with complex and differing shape, instillings confidence it captures the essential physics of dynamic AFM on soft materials.

\clearpage
\appendix*
\section{Methods}

\subsection{Measurement details}
Each measurement (one pixel of the scan) requires 2 ms, corresponding to a frequency spacing in the comb of intermodulation products of 500 Hz, also the measurement bandwidth. To enhance the signal-to-noise ratio, we average the measured intermodulation spectra over several neighboring pixels which show the same type of response, thereby using spatial correlation to reduce the noise.  This averaging results in smoother experimental force quadrature curves without reducing the scan speed.  

\subsection{Equations of motion}
For numerical simulation of the tip and surface motion it is convenient to introduce a non-dimensional time $u = \omega_0 t$, where $f_0 = \omega_0/2\pi$ is the resonant frequency of the cantilever in Hz.  With this scaling the relaxation times become dimensionless parameters, $u_\text{s} = \omega_0 \tau_\text{s}$ and $u_\text{v} = \omega_0 \tau_\text{v}$. Equations \eqref{eq:cant} and \eqref{eq:surf}  then describe a three-dimensional dynamical system with state variables $d$ , $d'$ and $d_\text{s}$.  For the case $s<0$ this system reads,
\begin{subequations}
    \begin{align}
        \frac{\mathrm{d}}{\mathrm{d} u} d   =& d' \label{eq:dp1}  \\
        \frac{\mathrm{d}}{\mathrm{d} u} d'  =& - \frac{1}{Q} d' - d + \frac{F_\text{d}}{k} - \frac{F_\text{ad}}{k} \nonumber\\
        &-\frac{k_v}{k}\left(s + u_\text{v}s'\right) \label{eq:dpp1}\\
        \frac{\mathrm{d}}{\mathrm{d} u} d_s =& d_s' =  - \frac{d_s}{u_\text{s}} + \frac{k_v}{u_\text{s} k_s} \nonumber\\
        &\times \left(\frac{F_\mathrm{ad}}{k_v} + s + u_\text{v}(d'-d_s')\right) \label{eq:dsp1}
    \end{align}
\end{subequations}
Since the right hand side of eq.~\eqref{eq:dpp1} depends on $s' = d' - d_s'$ at each time step, we must first solve eq.~\eqref{eq:dsp1} for $d_s'$ and then substitute it in to the right hand side of eq.~\eqref{eq:dpp1}.  In terms of the impact-rheology factor $R=\frac{u_\text{s}}{u_\text{v}}$ and the stiffness ratio $K=k_s/k_v$ we can write eq.~\eqref{eq:dpp1} as:
\begin{equation}
        \frac{\mathrm{d}}{\mathrm{d} u} d_s =  \frac{\eta_\text{n}}{\eta_\text{n} + \eta_\text{s}}\left[- \frac{d_s}{u_\text{s}} + \frac{1}{u_\text{s} K}\times \left(\frac{F_{ad}}{k_v} + s + u_\text{v}d'\right)\right] \label{eq:dsp2}
\end{equation}
The prefactor $\frac{1}{1+RK}$ in front of eq. \eqref{eq:dsp2} describes the strength of the dynamic coupling between tip and surface, giving a measure of how much surface motion one can expect.

\subsection{Numerical integration and optimization}

Numerical integration of the model is performed with the package CVODE, part of the SUNDIALS suite of nonlinear solvers~\cite{SUNDIALS}.  This integrator has adaptive time-stepping.  Because the dynamical system is coded in {\em C}, simulation time is reduced by a factor of 100 in comparison with coding in higher-level languages such as \textit{MATLAB} or \textit{Python}.  This speed up is of great importance as many integrations are required when iterating to find the optimal parameters.  When performing the numerical optimization, the initial conditions for $d$ and $d'$ are reset to their measured values for each iteration,  whereas the initial condition for $d_\text{s}$ is set to zero on the first iteration, and then estimated from the previous integration for all subsequent iterations.  

We use the Python library \textit{scipy.optimize.leastsq}, which is a wrapper for MINPACK's implementation of the Levenberg-Marquardt algorithm.  The algorithm performs a least-square minimization of the array of residuals $r = \{ \mathrm{real}(\hat{F}_{exp} - \hat{F}_{sim}) , \mathrm{imag}(\hat{F}_{exp} - \hat{F}_{sim}) \}$ where the $\hat{F}(\omega)$ are experimental and simulated complex force amplitudes at 40 frequencies.  The method finds a local minimum, starting from good initial parameter values determined by trial and error.

\subsection{Background force compensation}
When the cantilever is oscillating above a surface we often observe significant background forces, not resulting from tip-sample interaction but rather acting over the entire cantilever body.  The origin of these background forces might be \textit{e.g.} long-range electrostatic force or hydrodynamic squeeze-film damping.  In order to deduce the tip-sample force we need to remove this background interaction.  The procedure we use for removing any linear background force is described in a previous publication~\cite{Borgani:BackgroundForces:2017}. 

\subsection{Calibration}
Cantilever parameters are determined by the non-invasive thermal calibration method described by Higgins {\em et al.}~\cite{Higins:NoninvasiveCalibration:06}, which combines  the fluctuation dissipation theorem with Sader's method based on analysis of hydrodynamic damping~\cite{Sader:ArbitraryShape:2012}.  With this approach one can extract the three parameters of the cantilever transfer function, $k$, $f_0$ and $Q$, as well as the inverse responsivity of the optical detector used to measure cantilever deflection $\alpha^{-1}$[nm/V], all  from one measurement of the thermal Brownian motion of the cantilever near resonance.  The method is encapsulated in the recently launched Global Calibration Initiative (GCI) \cite{Sader:GCI:2016}, where a thermal noise measurement of $f_0$ and $Q$ can be used to get $k$, based on a single hydrodynamic constant determined from averaging over the measurements of may users on the same type of cantilever.  We take our measured $f_0$ and $Q$ and use the GCI to determine $k_\text{Sader}$, allowing us to then determine $\alpha^{-1}$.  Note that error in the calibration of $k$ and $\alpha^{-1}$ result in a re-scaling of the force and amplitude axes respectively, which does not change the general shape of the force quadrature curves.

\begin{acknowledgments}
We acknowledge S. Borysov, C. A. van Edysen, J. \mbox{Wettlaufer} and Xiao-min Lin for helpful discussions, and J. Jureller for help with the measurements. KTH authors are grateful for financial support from the Swedish Research Council (VR), the Olle Engkvist Foundation and the Knut and Allice Wallenberg Foundation. P.L. is FRS-FNRS Senior Research Associate.  Research in Mons was supported by the European Commission and Region Wallonne FEDER program, the Science Policy Office of the Belgian Federal Government (BELSPO-PAI VII/5), and the FRS-FNRS PDR Project ECOSTOFLEX.  Y.W. and H.M.J. acknowledge support from the Office of Naval Research through grant N00014-17-1-2342 and through the Chicago MRSEC under NSF DMR-1420709.
\end{acknowledgments}


\begin{thebibliography}{10}
\bibitem{RheologyImpactFactor}
Not to be confused with the journal Rheology, which has an impact-factor.
\end{thebibliography}

\end{document}